\documentclass{elsart}
\usepackage{amsfonts}
\usepackage{amssymb}
\usepackage{amsbsy}

\begin{document}

\begin{frontmatter}

\title{Mimicking a Kerrlike medium in 
the dispersive regime of  
second-harmonic generation} 

\author{A. B. Klimov}
\address{Departamento de F\'{\i}sica, 
Universidad de Guadalajara,  
Revoluci\'on 1500, 
44420~Guadalajara, Jalisco, 
Mexico} 

\author{L. L. S\'anchez-Soto and J. Delgado} 
\address{Departamento de \'Optica, 
Facultad de Ciencias F\'{\i}sicas, 
Universidad Complutense, 
28040 Madrid, Spain}

\maketitle

\begin{abstract}
We find an effective Hamiltonian describing the process 
of  second-harmonic generation in the far-off resonant 
limit. We show that the dynamics of  the fundamental 
mode is governed by a Kerrlike Hamiltonian. Some  
dynamical consequences are examined.

\textit{PACS number(s): 03.65.Bz,  42.50.Ct, 42.50.Vk, 42.65.Ky}

\textit{Section: Quantum Optics}
\end{abstract}

\end{frontmatter}

\section{Introduction}

The superposition principle plays an essential role in 
understanding the conceptual foundations of quantum 
mechanics. Stated in physical terms, it assures that 
we cannot speak of an objective state of a system 
prior to a measurement. In fact, one should admit 
that the system may be described as a superposition 
of macroscopically (or mesoscopically) distinct states, 
often known as Schr\"{o}dinger cat states~\cite{Sch83}.

Therefore, the methods for the generation of such superposition 
states are of fundamental interest. Among other systems, vibrating 
molecules~\cite{Jan94} or crystals~\cite{Wal95}, trapped 
ions~\cite{ion}, and Bose condensates~\cite{Cir98} have 
been proposed as candidates for doing the task. In fact, 
mesoscopic Schr\"{o}dinger cat states for the vibrational
motion of ion traps have been experimental realized~\cite{trapp}.

The issue of generating optical Schr\"{o}dinger cats (namely, 
superpositions of distinguishable states of the electromagnetic field) 
has attracted as well a great deal of attention~\cite{Ocats}. 
For example,  Brune \textit{et al}~\cite{QED} have shown that 
the conditional  measurement on atoms exiting a high-$Q$ cavity 
may force the radiation inside the cavity to be in a catlike
state. The same basic concept of conditional measurement 
has also been suggested for entangled states~\cite{cment}, 
nonlinear birefringence~\cite{Mec87}, or self-Kerr phase 
modulation~\cite{Yur86}. The state collapse due to 
photodetection has also been suggested as a source of cat 
states~\cite{Ban96}.

The Kerr effect provides a nonlinearity of particular interest for
generating field cat states~\cite{Kerr}. Apart from their intrinsic
simplicity, Kerr-based schemes have the specific advantage of not 
relying on conditional measurements. However, the realistic values 
of Kerr coefficients are quite small, thus requiring a large 
interaction length. Then, losses become significant and may 
destroy the very delicate quantum superpositions. In short, 
although very appealing from the physical viewpoint, Kerr
schemes are not generally considered to be realistic. In spite of 
this belief, there has been a very recent proposal of generating a 
Kerr \textit{kitten} that serves as a quantum seed for suitable 
amplification~\cite{Par99}, that could become fully feasible.

On the other hand, it should be noted that perhaps the simplest 
nonlinear optical process is second-harmonic generation, which 
in addition  exhibits a rich spectrum of nonclassical features. 
Considerable  attention has been paid to issues such as photon 
antibunching,  squeezing, or collapses and revivals~\cite{ncshg}. 
However,  the generation of Schr\"{o}dinger cat states has gone
almost unnoticed. Perhaps, a relevant exception is the work of 
Nikitin and Masalov~\cite{NiMa91}, showing that, at resonance, 
the quantum state of the fundamental mode evolves into a 
superposition of two macroscopically distinguishable states. 
This point may be considered as well rather unrealistic, 
because the stringent experimental constraints of having
perfect phase matching and low decoherence could be rather 
difficult to attain.

In this paper we pursue the dispersive limit of second-harmonic 
generation, which seems to be almost ignored in the literature. 
At this respect, our main result is to show that, for detunings 
large enough, the dynamics can be described in terms of an 
effective Hamiltonian essentially identical to that governing 
the propagation of the fundamental mode in a Kerr medium.

Apart from its intrinsic interest, such an identification could be 
more than an academic curiosity: second-harmonic generation is, 
for a variety of reasons~\cite{Whi00}, more robust than the Kerr 
effect as for noise-limiting factors. In consequence, this scheme 
could be an experimentally feasible proposal to generate the 
yearned optical cat states.

\section{Dynamics of second-harmonic generation in the 
dispersive limit}

Second-harmonic generation is usually described (without 
assuming perfect resonance) by the following model Hamiltonian 
(in units $\hbar =1$)  
\begin{equation}
H=H_0+H_{\mathrm{int}} = \omega_a a^\dagger a + 
\omega_b b^\dagger b + g(a^2 b^\dagger +{a^\dagger }^2 b),  
\label{Hshg}
\end{equation}
where $a$ ($a^\dagger$) and $b$ ($b^\dagger$) are 
the annihilation (creation) operators of the fundamental mode 
of frequency $\omega_a$ and of the second-harmonic mode 
of frequency $\omega_b$, respectively. When perfect 
matching conditions occurs, they satisfy $\omega_b= 2 
\omega_a$. The constant $g$, describing phenomenologically 
the coupling between both modes, can be always chosen as real.

Classically the above problem admits an exact solution, leading 
to the possibility of the energy being transferred completely into 
the second-harmonic mode~\cite{Blo72}. Unfortunately, the 
corresponding quantum dynamics is a touchy business, though 
different algebraic, semiclassical, and numerical approaches 
have been developed (for pertinent literature on these 
approximations see Ref.~\cite{quant}).

By noting that the Hamiltonian (\ref{Hshg}) admits the 
constant of motion  
\begin{equation}
N = a^\dagger a + 2 b^\dagger b ,  \label{N}
\end{equation}
it can be recast in the following simple form 
\begin{eqnarray}  
\label{Hint}
H_0 & = & \frac{\omega_b + \omega_a}{3} N ,  
\nonumber \\
& & \\
H_{\mathrm{int}} & = & \frac{\Delta}{3} 
( b^\dagger b - a^\dagger a ) + 
g (a^2 b^\dagger +{a^\dagger}^2 b ) ,  \nonumber
\end{eqnarray}
where 
\begin{equation}
\Delta = \omega_b - 2 \omega_a
\end{equation}
is the detuning.

It is straightforward to check that 
\begin{equation}
[H_0, H_{\mathrm{int}} ] = 0 ,
\end{equation}
so both $H_0$ and $H_{\mathrm{int}}$ are constants 
of  motion. Since $H_0$ determines the total energy 
stored in both modes, which is conserved, we can
factor out $\exp(- i H_0 t)$ from the evolution operator 
and drop it altogether.

In order to examine the quantum evolution we shall use 
a standard numerical approach developed by Tana\'{s} 
and coworkers~\cite{Tan91} starting from the original 
one of Walls and Barakat~\cite{Wal70}. First, we consider 
the case of perfect resonance and we take the typical 
experimental conditions; namely, the fundamental mode 
in a coherent state with an average number of photons 
$\bar{n}_a$ and the harmonic mode in the vacuum.

For definiteness, we shall analyze the quantum state by 
computing the $Q$ function. In figure~1 we have plotted 
a gray-scale contour map of this distribution $Q(\alpha)$ 
for the fundamental mode with $\bar{n}_a = 10$, and
for different values of the dimensionless time $\tau = 
g t \sqrt{2 \bar{n}_a}$. Results are in complete agreement 
with the previous work by Nikitin and Masalov~\cite{NiMa91}: 
at the beginning of the interaction $(\tau < 2)$ the uncertainty 
contour turns to an ellipse, whose long axis is perpendicular to
the field vector (so the state is squeezed in amplitude). Further
interaction $(\tau \simeq 4)$ results in a bifurcation of the $Q$
distribution and the formation of a superposition of two 
macroscopically distinguishable states. Subsequent development 
$(\tau > 5)$ destroys this state as well.

Next, we consider the case when some mismatch is present. 
The previous numerical procedure is still valid, but one must 
take care of the diagonal extra terms introduced by the detuning. 
To check how the detuning affects the cats, in figure~2 we have 
plotted the contours of the $Q$ function at $\tau =4$, when the 
cat should appear, for different values of the dimensionless 
detuning $\Delta /g$. One easily notices that the cat becomes 
more and more blurred as the detuning increases, 
until $\Delta /g$ is so large and the process is so 
inefficient that the fundamental mode does not feel any 
appreciable interaction and the cat is lost. Moreover, we
have found no numerical evidence of any further cat for 
other times around $\tau =4$.

This is a reasonable result from the physical viewpoint, since 
one could expect some kind of continuity. However, it differs 
from the numerical computations of Nikitin and 
Masalov~\cite{NiMa91}, who find that the cat disappears 
as soon as some detuning is present, irrespectively the smaller
this detuning is. The reason is the way they deal with 
the off-resonant situation: following a classical argument, they 
assume that the presence of detuning translates into the fact that 
the coupling constant has the time dependence $g=|g|
\exp[(n_a - n_b) \omega t]$, where $n_a$ and $n_b$ are the 
refractive indices of the medium at the frequencies 
$\omega_a$ and $\omega_b$. In such a way, the interaction 
Hamiltonian $H_{\mathrm{int}}$ becomes time dependent 
and the numerical method becomes rather complicated, 
at difference of what happens with our very 
simple model.

From now on, we shall be interested in the dispersive limit 
of the model governed by the Hamiltonian (\ref{Hshg}). 
In other words, we shall consider the situation when 
\begin{equation}
\label{displim}
|\Delta | \gg  g (\bar{n}_a + 1 ) (\bar{n}_b + 1),
\end{equation}
where $\bar{n}_a$ and $\bar{n}_b$ denote the average 
photon numbers in the modes $a$ and $b$, respectively. 
It is worth emphasizing that this is a limit physically 
realizable in practice~\cite{Bru96}. Now, the essential
point is that $g/\Delta $ can be considered as a small 
parameter and, according to the method developed in 
Ref.~\cite{Kli00}, we can apply a unitary transformation 
to the interaction Hamiltonian~(\ref{Hint}) 
\begin{equation}
H_{\mathrm{eff}} = U H_{\mathrm{int}}U^{\dagger },  
\label{H1}
\end{equation}
where 
\begin{equation}
U = \exp \left [ \frac{g}{\Delta }
(a^2 b^\dagger - {a^\dagger}^{2}b ) 
\right] .  \label{U}
\end{equation}
To gain a deeper insight into the physical ideas underlying 
the method we can introduce the operators 
\begin{eqnarray}
& X_+ = b^\dagger a^2 , \qquad 
X_- ={a^\dagger}^2 b, & \nonumber \\
&& \\
&X_3 = \frac{1}{3}( b^\dagger b - a^\dagger a) ,  &
\nonumber
\end{eqnarray}
that satisfy the commutation relations 
\begin{eqnarray}
\left[ X_3, X_\pm \right] & = & \pm X_\pm ,  
\nonumber \\
& & \\
\left [ X_+, X_- \right] &=& P(X_3),  \nonumber
\end{eqnarray}
where $P$ refers to a quadratic polynomial function of 
the diagonal operator  $X_3$. Then, while the first 
relation is the same as the corresponding one for the 
standard su(2) algebra, the second one defines what is 
called a  nonlinear or polynomial deformation of 
su(2)~\cite{Kli00}.  From this viewpoint, the 
transformation~(\ref{U}) can be recast as
\begin{equation}
U=\exp \left[ \frac{g}{\Delta }(X_+ - X_- ) \right] . 
\end{equation}
For su(2) operators this would be simply a rotation. 
Here, it represents a small nonlinear rotation with the 
parameters chosen in such a way so as to cancel 
[up to order $(g/\Delta )^2]$ the nondiagonal terms 
appearing in the transformed Hamiltonian. Moreover, 
its  unitary character ensures that  
$H_{\mathrm{eff}}$ has the same spectrum as 
$H_{\mathrm{int}}$ (so they are physically equivalent). 
In consequence, by expanding Eq.~(\ref{H1}) in a power 
series and keeping terms up to the order $(g/\Delta )^2$, 
we finally get 
\begin{equation}
H_{\mathrm{eff}}=\frac{\Delta }{3}
(b^\dagger b - a^\dagger a)-
\lambda \left[ 4b^\dagger b a^\dagger a-
(a^\dagger a)^2 \right] ,  \label{Heff}
\end{equation}
with 
\begin{equation}
\lambda =\frac{g^{2}}{\Delta }.
\end{equation}

The effective Hamiltonian~(\ref{Heff}) describes 
the dispersive evolution of the fields, but the 
essential point is that it is diagonal, which implies
that there is no population transfer between the 
modes (as it is expected in the far-off resonant limit). 
The first term in Eq.~(\ref{Heff}) does not affect 
the dynamics and just leads to a rapid oscillation of the 
wavefunction. It  is worth noting that the small rotation 
generated by the transformation~(\ref{U}) just adds 
small corrections of order $g/\Delta $ to the physical
observables and, thus, the state vector need not to be 
transformed.

Now, it is easy to find evolution of the density matrix 
of the fundamental mode. When the harmonic mode 
is initially in the vacuum, the two terms containing 
$b^\dagger b$ in Eq.~(\ref{Heff}) do not contribute. 
In addition,  the linear term in $a^{\dagger }a$ lead just 
to a  $c$-number phase shift and can be also omitted. Then, 
the effective Hamiltonian reduces to 
\begin{equation}
H_{K}=\lambda (a^\dagger a)^2,  
\label{HK}
\end{equation}
which is nothing but the interaction Hamiltonian that 
governs the state evolution of the single-mode field $a$ 
in a Kerr medium. Then, the evolution is factorized all 
the times (in the frame of this approximation) and, in 
consequence, starting from a coherent state in mode 
$a$, one will recover $M$ copies of this initial state at 
times $\lambda t=\pi /M$~\cite{Kerr}. In particular, 
for $\lambda t=\pi /2$ the density matrix of the 
fundamental mode becomes
\begin{equation}
\rho_a (\pi /2\lambda ) = |\psi _{c}\rangle_a  \ {}_a\langle \psi _{c}| , 
\end{equation}
where $|\psi _{c} \rangle_a$ is a superposition of two 
coherent states with opposite phases; i.e.,
\begin{equation}
|\psi _{c}\rangle_a = \frac{1}{\sqrt{2}}
\left[ e^{i\pi /4} |\alpha \rangle_a +
e^{-i \pi /4} |-\alpha \rangle_a \right] . 
\end{equation}
To check these predictions we have numerically calculated 
the $Q$ function from the exact Hamiltonian (\ref{Hint}) 
at the time $t=\pi /(2\lambda )$. The initial coherent state in 
mode $a$ has $\bar{n}_a=10$ and we have assumed a 
detuning $\Delta /g=50$, which is enough to guarantee the 
validity of our approximation~(\ref{displim}). In figure 3 we 
have plotted the results of the computation, showing an apparent excellent 
agreement with the expected cat. However, the $Q$ function 
by itself is not very suitable to distinguish between coherent 
and incoherent superpositions. To further confirm the
coherent behavior we have considered the fidelity
\begin{equation}
\mathcal{F}(t)=  {}_a \langle \psi _{c}| \rho (t) |
\psi _{c}\rangle_a  
\end{equation}
between the ideal two-component cat density matrix and 
the state coming from realistic evolution. In figure 4 we 
have plotted the resulting fidelity as a function of the rescaled time 
$g t$ for $\Delta /g=50.$ We can see the presence of small peaks 
corresponding to the presence of $M-$component catlike states,
as well as the strong peak at the time $\lambda t = \pi/2$ 
predicted by our theory  when $M=2$.

When the initial state of the harmonic mode is in a superposition 
of number states instead of the vacuum, the dynamics of 
the fundamental mode is drastically affected. For simplicity, 
we shall take an initial coherent state in both modes, 
namely, $|\alpha \rangle _a \otimes |\beta \rangle_b .$ 
Then, we obtain for the density matrix of the fundamental 
mode the following expression 
\begin{eqnarray}
\rho_a (t) &  = &  e^{- |\alpha|^2} 
\sum_{n,m}^\infty  
e^{it\lambda (n^2 - m^2)} 
\exp \left[ |\beta|^2 \left ( 1-e^{-4i\lambda t(n-m)}\right) \right] 
\nonumber \\
& \times & \frac{\alpha^n {\alpha^*}^m}{\sqrt{n!m!}}
|n\rangle_a \  {}_a\langle m|.
\end{eqnarray}
The entanglement between modes $a$ and $b$ is evident, as
well as the fact that only when $\lambda t=\pi /2$ they become 
disentangled and two copies of the initial coherent state in 
the  fundamental mode will be created. Indeed, the density 
matrix of the total system at such time is  
\begin{equation}
\rho (\pi /2\lambda ) = |\beta \rangle_b \ {}_b\langle \beta |
\otimes |\psi_c \rangle_a  \ {}_a \langle \psi_{c}|,
\end{equation}
and only a two-component cat can be produced in this case,
although with lesser fidelity than for the vacuum. On 
the contrary,  when $\lambda t= \pi$, the initial coherent state 
is reconstructed.

Finally, some comments concerning the physical feasibility 
of the proposed scheme seem in order. Of course, a simple 
comparison of the parameters in figures 1 and 3 clearly 
shows that the times required for the appearance of the cat in 
the dispersive limit are an order of magnitude larger than in 
the usual case of no detuning. The advantage of this dispersive 
limit is twofold: one does not need to take care about 
phase-matching considerations and, more important, if  the 
process takes place in a cavity, one could expect from previous 
results, e.g. in cavity quantum electrodynamics~\cite{Bru96}, that 
the quantum features of the state may be undestroyed for 
large interaction times. In addition, the width of the peak
appearing in the fidelity is of the order of 5 in units of $ g t$,
while the cat appearing in figure 1 disappears in much shorter
times. Thus, the required time resolution in this dispersive limit 
is lesser than the one needed for the perfect frequency-matching 
case. In spite of these advantages, the required times are still 
so long that other  decoherence factors could affect the final 
results.

The effective Hamiltonian (\ref{Heff}) has the virtue of 
simplifying calculations that otherwise would be rather 
involved. To cite only an example, the squeezing 
properties of the fundamental mode are also affected
by the initial state of the harmonic mode. For initial 
coherent states in both modes $|\alpha \rangle_a 
\otimes |\beta \rangle_b $ (for simplicity, we
take $\alpha $ and $\beta $ as real numbers) 
we easily get for the variance $\sigma_x^2=
\langle x^2 \rangle - \langle x \rangle^2$ of the 
quadrature $x = (a + a^\dagger )/ \sqrt{2}$ 
\begin{eqnarray}
\sigma_x^2 (t) & = & \frac{1}{2} + 
\alpha^2 (e^{-8 z^2 T^2} - e^{- 4 z^2 T^2} ) 
\cos \Omega T  \nonumber \\
& - & 2 \alpha^2 T^2 
( 4e^{-8 z^2 T^2 }- e^{- 4 z^2 T^2} ) 
\cos \Omega T  
\nonumber \\
& - & 2 \alpha^2 T ( 2e^{-8 z^2 T^2} - 
e^{-4 z^2 T^2} ) \sin \Omega T 
\nonumber \\
& + & \alpha^2 
( 1 - e^{- 4 z^2 T^2} ) \cos \Omega T,
\end{eqnarray}
where $T= \lambda t \ll 1$, $z^2 = \alpha^2 + 4 \beta^2$, 
and  $\Omega = 4 \alpha^2-8 \beta^2$. Then, for 
sufficiently large values of $\beta^2$ squeezing in 
the fundamental mode disappears, according to previous
numerical studies~\cite{NiMa91}.

One could be tempted to extend the method to other 
nonlinear optical process in order to get effective 
Hamiltonians that are diagonal in the desired basis. 
In fact, the program can be carried out. For example, 
for third-harmonic generation one obtains, after repeating 
the same steps as before, 
\begin{equation}
H_{\mathrm{eff}} \simeq \lambda \left \{ 
9 b^\dagger b \left [ (a^\dagger a)^2 + 
a^\dagger a \right ] - (a^\dagger a)^3 - 
6 (a^\dagger a)^2 \right \} ,
\end{equation}
but now the final result contains a cubic nonlinearity and 
the problem is not facilitated at all.

\section{Concluding remarks}

What we expect to have accomplished in this paper is 
to present a simple model for the description of the 
dispersive limit of second-harmonic generation in terms 
of an effective Hamiltonian that is diagonal and contains 
Kerrlike nonlinearities. Of course, in the practice 
other off-resonant processes could  also contribute  to 
the third-order nonlinearity. However, our effective-Hamiltonian 
approach can explain (al least qualitatively) some unexpected
and fascinating behaviors of the so-far unnoticed
dispersive limit.  

Of course, it is well known that when some interaction 
energy is off-resonant then, in a higher order of  perturbation 
theory an effective interaction  energy can be 
constructed that gives rise to resonant processes. 
At this respect, in Ref.~\cite{Kli00} it has
been shown how the method of small rotations
can be used to tailor resonant interactions
from nonresonant processes, just as demonstrated in
this paper for the relevant case of the second-harmonic 
generation. 

Finally, it is worth noting as well that 
the Hamiltonian~(\ref{Heff})  is similar to that obtained 
in Ref.~\cite{dicke} for the Dicke model in the large 
detuning limit  (when the effective Hamiltonian was 
proportional to $S_z^2$).

\newpage

\begin{figure}
\caption{Gray-level contour plot of the 
quasiprobability distribution $Q(\alpha)$ 
of the fundamental mode (with perfect frequency matching)  
at different times $\tau = g t \sqrt{2 \bar{n}_a}$. 
Initially, the fundamental mode is in a coherent state with 
$\bar{n}_a = 10$ and the harmonic mode is in the vacuum. 
In all the cases in the horizontal  axis we plot 
$\mathrm{Re} \ \alpha$ and in the vertical we plot 
$\mathrm{Im} \ \alpha$.}
\end{figure}

\begin{figure}
\caption{Gray-level contour plot of the quasiprobability distribution 
$Q(\alpha)$ of the fundamental mode for the same initial state as in 
figure 1 calculated for $\tau = 4$ at different values of $\Delta/g$.}
\end{figure}

\begin{figure}
\caption{$Q$ function and the corresponding contour plot 
of the fundamental mode (initially, the fundamental mode 
is in a coherent state with $\bar{n}_a = 10$ and the harmonic 
mode is in the vacuum)  with a detuning of $\Delta/g = 50$. 
The function has been calculated calculated from the exact
interaction Hamiltonian~(3) at $\lambda t=\pi /2$. }
\end{figure}

\begin{figure}
\caption{Fidelity of the fundamental mode respect to the ideal 
cat state as a function of the rescaled time $g t$. The state 
evolution has been calculated from the exact interaction 
Hamiltonian~(3). The initial state is as in figure 3 with 
the same detuning $\Delta/g = 50$. The main peak appears 
at the time $\lambda t=\pi /2$ predicted by the theory.}
\end{figure}


\begin{thebibliography}{99}
\bibitem{Sch83}  
E. Schr\"{o}dinger, in \textit{Quantum Theory of Measurement}, 
edited by J. A. Wheeler and W. H. Zurek 
(Princeton University Press, Princeton, NJ) p. 152.

\bibitem{Jan94}  
J. Janzky, A. V. Vinogradov, T. Kobayashy, Z. Kis, 
Phys. Rev. A \textbf{50}, 1777 (1994)

\bibitem{Wal95}  
I. A. Walmsley, M. G. Raymer, 
Phys. Rev. A \textbf{52}, 681 (1995)

\bibitem{ion}  
J. J. Slosser, P. Meystre, E. M. Wright, 
Opt. Lett. \textbf{15}, 233 (1990)

R. L. de Matos-Filho, W. Vogel, 
Phys. Rev. Lett. \textbf{76}, 608 (1996)

M. M. Nieto, 
Phys. Lett. A \textbf{219}, 180 (1996)

\bibitem{Cir98}  
J. I. Cirac, M. Lewenstein, K. Molmer, P. Zoller, 
Phys. Rev. A \textbf{57}, 1208 (1998)

\bibitem{trapp}  
C. Monroe, D. M. Meekhof, B. E. King, D. J. Wineland,
Science \textbf{272}, 1131 (1996)

D. M. Meekhof, C. Monroe, B. E. King, W. M. Itano, D. J. Wineland, 
Phys. Rev. Lett. \textbf{76}, 1796 (1996)

D. Leibfried, D. M. Meekhof, B. E. King, C. Monroe, 
W. M. Itano, D. J. Wineland, 
Phys. Rev. Lett. \textbf{77}, 4281 (1996)

D. J. Wineland, C. Monroe, D. M. Meekhof, D. Leibfried, 
W. M. Itano, J. C. Bergquist, D. Berkeland, J. J. Bollinger, J. Miller, 
Proc. Roy. Soc. A  \textbf{454}, 411 (1998)

\bibitem{Ocats}  
V. V. Dodonov, I. A. Malkin, V. I. Man'ko, 
Physica \textbf{72}, 4281 (1974)

M. Hillery, 
Phys. Rev. A \textbf{36}, 3796 (1987)

G. C. Gerry, 
Opt. Commun. \textbf{63}, 278 (1987)

R. Lynch, 
Opt. Commun. \textbf{67}, 67 (1988)

W. Schleich, M. Pernigo, F. Le Kien, 
Phys. Rev. A \textbf{44}, 2172 (1991)

V. Bu\v {z}ek, H. Moya-Cessa, P. L. Knight, S. J. D. Phoenix, 
Phys. Rev. A  \textbf{45}, 8190 (1992)

\bibitem{QED}  
M. Brune, S. Haroche, J. M. Raimond, L. Davidovich, N. Zagury, 
Phys. Rev. A \textbf{45}, 5193 (1992)

L. Davidovich, M. Brune, J. M. Raimond, S. Haroche, 
Phys. Rev. A \textbf{53}, 1295 (1996)

M. Brune, E. Hagley, J. Dreyer, X. Ma\^{\i }tre, A. Maali, 
C. Wunderlich, J. M. Raimond, S. Haroche, 
Phys. Rev. Lett. \textbf{77}, 4887 (1996)

\bibitem{cment}  
S. Song, C. M. Caves, B. Yurke, 
Phys. Rev. A \textbf{41}, 5261 (1990)

B. Yurke, W. P. Schleich, D. F. Walls, 
Phys. Rev. A \textbf{42}, 1703 (1990)

P. Tombesi, D. Vitali, 
Phys. Rev. Lett. \textbf{77}, 411 (1996)

M. Dakna, T. Anhut, T. Opatrny, L. Knoll, D. G. Welsch, 
Phys. Rev. A \textbf{55}, 3184 (1997)

\bibitem{Mec87}  
A. Mecozzi, P. Tombesi, 
Phys. Rev. Lett. \textbf{58}, 1055 (1987)

\bibitem{Yur86}  
B. Yurke, D. Stoler, Phys. Rev. Lett. \textbf{57}, 13 (1986)

S. D. Du, S. Gong, Z. Z. Xu, L. W. Zhou, C. D. Gong, 
Opt. Commun. \textbf{138}, 193 (1997)

\bibitem{Ban96}  
M. Ban, 
J. Mod. Opt. \textbf{43}, 1281 (1996)

M. Ban, 
Phys. Lett. A \textbf{233}, 284 (1997)

A. Luis, L. L. S\'{a}nchez-Soto, 
Phys. Lett. A \textbf{244}, 211 (1998)

\bibitem{Kerr}  
M. Kitagawa, Y. Yamamoto, 
Phys. Rev. A \textbf{34}, 3974 (1986)

G. S. Milburn, 
Phys. Rev. A \textbf{33}, 674 (1986)

R. Tana\'{s}, Ts. Gantsog, A. Miranowicz, S. Kielich, 
J. Opt. Soc. Am. B  \textbf{8}, 1576 (1991)

R. Tara, G. S. Agarwal, S. Chaturvedi, 
Phys. Rev. A \textbf{47}, 5024 (1993)

G. V. Varada, G. S. Agarwal, 
Phys. Rev. A \textbf{48}, 4062 (1993)

\bibitem{Par99}  
M. G. A. Paris, 
J. Opt. B: Quantum Semiclass. Opt. \textbf{1}, 662 (1999)

\bibitem{ncshg}  
M. Kozierovski, R. Tana\'{s}, 
Opt. Commun. \textbf{21}, 229 (1977)

L. Mandel, 
Opt. Commun. \textbf{42}, 437 (1982)

L. Wu, H. J. Kimble, J. L. Hall, H. Wu, 
Phys. Rev. Lett. \textbf{57}, 2520 (1986)

G. Drobn\'{y}, I. Jex, 
Phys. Rev. A \textbf{46}, 499 (1992)

G. Drobn\'{y}, I. Jex, V. Bu\v {z}ek, 
Phys. Rev. A \textbf{48}, 569 (1993)

\bibitem{NiMa91}  
S. P. Nikitin, A. V. Masalov,  
Quantum Opt. \textbf{3}, 105 (1991)

\bibitem{Whi00}  
A. G. White, P. K. Lam, D. E. McClelland, 
H-A. Bachor, W.J. Munro, 
J. Opt. B: Quantum Semiclass. Opt. \textbf{2}, 553 (2000)

\bibitem{Blo72}  
N. Bloembergen, \textit{Nonlinear Optics} 
(McGraw-Hill, New York, 1972)

\bibitem{quant}  
R. F. Alvarez-Estrada, A. G\'{o}mez-Nicola, 
L. L. S\'{a}nchez-Soto, A. Luis, 
J. Phys. A \textbf{28}, 3439 (1995)

G. Alvarez, R. F. Alvarez-Estrada, 
J. Phys. A \textbf{28}, 5767 (1995)

\bibitem{Tan91}  
R. Tana\'{s}, Ts. Gantsog, R. Zawodny, 
Quantum Opt. \textbf{3}, 221 (1991)

\bibitem{Wal70}  
D. F. Walls, R. Barakat, 
Phys. Rev. A \textbf{1}, 446 (1970)

\bibitem{Bru96}  
M. Brune, E. Hagley, J. Dreyer, X. Ma\^{i}tre, 
A. Maali, C. Wunderlich, J. M. Raimond, S. Haroche, 
Phys. Rev. Lett. \textbf{77}, 4887 (1996)

\bibitem{Kli00}  
A. B. Klimov, L. L. S\'{a}nchez-Soto, 
Phys. Rev. A \textbf{61}, 063802 (2000)

\bibitem{dicke}  
G. S. Agarwal, R.R.Puri, R.P.Singh, 
Phys. Rev. A \textbf{56}, 2249 (1997)
\end{thebibliography}
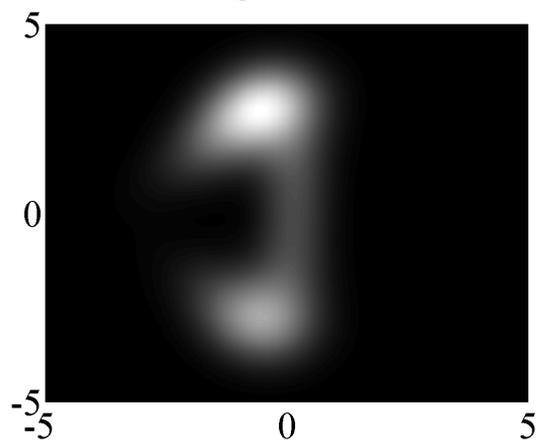
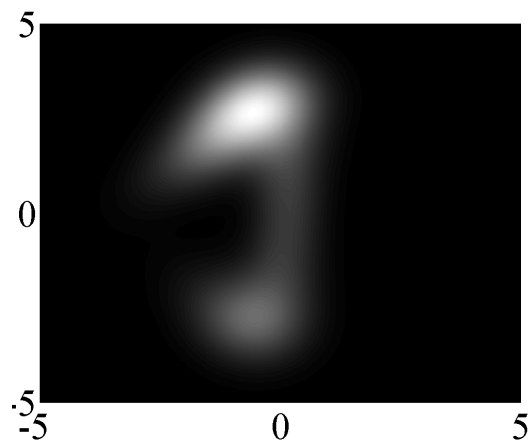
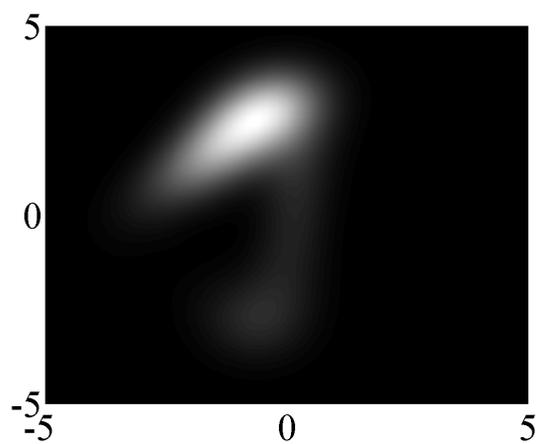
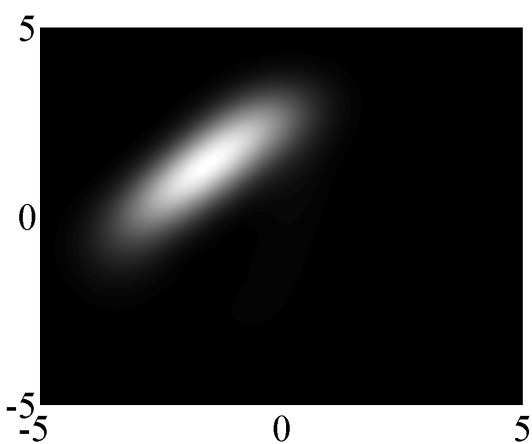
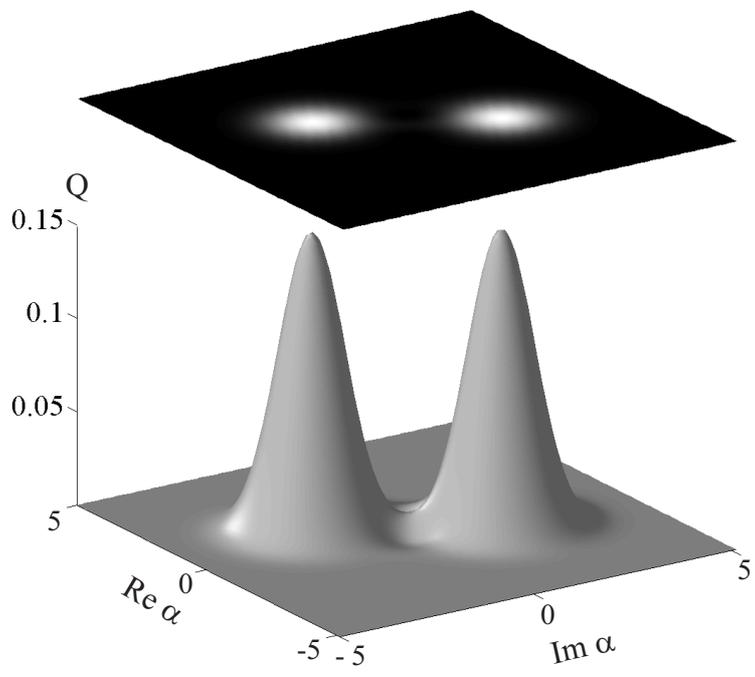
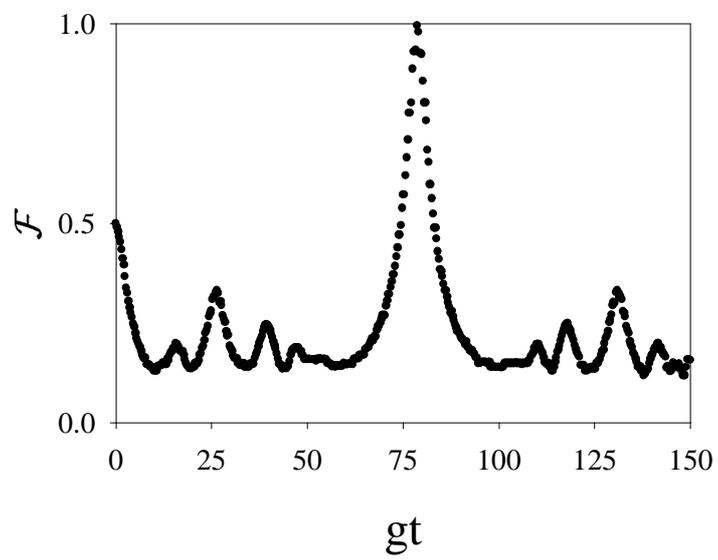
\end{document}